\documentclass[preprint]{aastex}
\usepackage{natbib}
\usepackage{graphicx}
\usepackage{subfigure}
\usepackage{amsmath,amssymb}
\begin{document}

\title{UV and EUV Emissions at the Flare Foot-points Observed by AIA}
\author{Jiong Qiu$^1$, Zoe Sturrock$^2$, Dana W. Longcope$^1$, James, A. Klimchuk$^3$, Wen-Juan, Liu$^1$}
\affil{1. Department of Physics, Montana State University, Bozeman MT 59717-3840, USA\\
2. Department of Applied Mathematics, University of St. Andrews, UK\\
3. NASA Goddard Space Flight Center, Greenbelt, MD 20771, USA}

%
%
%
\begin{abstract}
A solar flare is composed of impulsive energy release events by magnetic reconnection,
which forms and heats flare loops. Recent studies have revealed a {\em two-phase} 
evolution pattern of UV 1600\AA\ emission at the feet of these loops: a rapid 
pulse lasting for a few seconds to a few minutes, followed by a gradual decay on timescales of a 
few tens of minutes. Multiple band EUV observations by AIA further reveal very similar signatures. 
These two phases represent different but related signatures of an impulsive energy release in the corona.  
The rapid pulse is an immediate response of the lower atmosphere to an intense thermal conduction 
flux resulting from the sudden heating of the corona to high temperatures (we rule out energetic particles 
due to a lack of significant hard X-ray emission). The gradual phase is associated with the cooling of hot 
plasma that has been evaporated into the corona. The observed footpoint emission is again powered by thermal 
conduction (and enthalpy), but now during a period when approximate steady state conditions are established in the loop.
UV and EUV light curves of individual pixels may therefore be separated into contributions from two distinct physical 
mechanisms to shed light on the nature of energy transport in a flare.  We demonstrate this technique using coordinated, spatially 
resolved observations of UV and EUV emission from the footpoints of a C3.2 thermal flare.

\end{abstract}

\section{Introduction}
The source of flare energy is magnetic, but the ultimate form of flare energy 
output is radiation of all kinds: lines and continnua by thermal and 
non-thermal particles. The bulk of X-ray and EUV radiation is produced in 
flare coronal loops, whereas enhanced optical, UV, and hard X-ray emissions 
are usually observed at the foot-points of these flare loops. 
Flare plasmas and particles are magnetically confined to be able to communicate mainly
along the loop from its lower-atmosphere root to the corona. Therefore, various radiation 
signatures along a flare loop are coherently coupled by physics governing energetics and dynamics
of magnetized flare plasmas.

In general, we may separate the energy release process along a flare loop
into the heating phase and cooling phase. A series of hydrodynamic responses
take place in an impulsively heated flaring atmosphere. A downward heat flux or energetic particle
beam generates a localized pressure pulse that drives bi-directional flow: an evaporation 
upflow into the corona and a condensation downflow into the chromosphere \citep{Canfield1986}. 
Evaporation sends heated plasma into the corona, which then cools as it radiates in X-ray and EUV wavelengths. 
Note that thermal conduction cooling of the evaporated material drives further evaporation, but 
this is a quasi-steady process, different from the initial explosive evaporation.  
Finally, the late stage of cooling is dominated by radiation and involves a slow draining of the 
material back onto the chromosphere. This process of flare loop evolution is often demonstrated by observations showing 
that bulk X-ray and EUV emissions in the corona are delayed with respect to the impulsively 
rising hard X-ray, UV, and optical emissions from the lower atmosphere. 

Numerous spectroscopic observations have unravelled dynamics in the early
(heating) phase of the flare, showing upflows of up to a few hundred kilometers per second 
in hot lines formed at a few million degrees \citep{Antonucci1982, Doschek1992,
Culhane1992, Bentley1994}, as well as downflows of several tens of kilometers per second 
in the chromospheric H$\alpha$ line \citep{Ichimoto1984, Canfield1987a, Schmieder1987, Canfield1987b, Zarro1988,
Wuelser1989, Canfield1990a, Canfield1990b, Wuelser1994, Ding1995}. Spectroscopic observations by recent missions such as 
the Coronal Diagnostic Spectrometer \citep[CDS;][]{Harrison1995} and the EUV Imaging Spectrometer \citep[EIS;][]{Culhane2007} 
have also identified these dynamical phenomena in UV and EUV lines at the feet of flare loops, where sometimes hard X-ray sources are 
located \citep{Milligan2006a, Milligan2006b, Milligan2009, Watanabe2010, DelZanna2011, Graham2011}. 
It takes a rather short time for energy flux carried by either non-thermal particles or 
thermal conduction to reach the lower atmosphere and enhance UV and optical emission \citep{Fisher1985, Canfield1987}. 
Therefore, the impulsive and dynamic behavior of radiation at the lower atmosphere, usually ahead of siginificant
coronal emissions, are registered as prompt signatures of flare energy release.

On the other hand, during the cooling phase, observations of some stellar flares
have shown that emissions in a few optical and UV bands appear to decay rather 
gradually at a rate very similar to the timescale of coronal radiation \citep{Hawley1992, Hawley2003}. 
Similar behavior of UV light curves was observed in some flares by Solar Maximum Mission (SMM) \citep[e.g.][]{Cheng1987}.
With high-resolution observations by the Transition Region And Corona Explorer \citep[TRACE;][]{Handy1999}
and the Atmosphere Imaging Assembly \citep[AIA;][]{Lemen2012}, \citet{Qiu2010, Qiu2012, Cheng2012, Liu2013} 
have also found that the broadband 1600\AA\ UV emission from individual pixels (1 \arcsec\ by 1 \arcsec) 
exhibits two-phase evolution characterized by a rapid rise and a gradual decay. During the cooling phase, 
conductive flux continuously flows from the corona toward the lower atmosphere - the transition 
region and chromosphere, which cools off by radiation. It is therefore considered that the prolonged 
decay in the lower atmosphere emission is coupled with coronal evolution, and may serve as a coronal 
``pressure gauge'' \citep{Fisher1987, Griffiths1998, Hawley2003}.

Separating the radiative signatures from the footpoint {\em of a single loop} into 
two distinct physical contributions provides crucial observational constraints to 
flare models. \citet{Fisher1990}
modeled an observed solar flare with a heating rate assumed to have the same time 
profile of the observed hard X-ray light curve. Quite a few solar flare studies 
followed a similar approach using (spatially unresolved) hard X-ray light curves or energy flux converted from
spectral analysis to infer impulsive energy release rate in the flare loop \citep[e.g.][]{Raftery2009}.
Taking advantage of high resolution UV imaging observations, \citet{Qiu2012, Liu2013} recently
modeled heating of thousands of flare loops (with nominal cross-section of 1\arcsec\ by 1\arcsec) 
using heating rates inferred from the rise phase of the UV emission at the feet of these flare loops. 
Using UV signatures to build heating rates, these studies not only resolve heating in individual loops, but
are not confined to flares that have significant thick-target hard X-ray emissions. It should be noted 
that the subsequent decay of the UV emission at these same feet, which is considered to be governed by 
evolution of the overlying flaring corona, should depend on the heating history. Along this line, 
\citet{Liu2013} conducted modeling and analysis of an M8.0 flare, and computed UV emission in the cooling 
phase. The result has shown, for the first time, that the the computed UV 
emission is in good agreement with the observed UV flux, and both decay at the same rate. 

In this paper, we report UV and EUV observations of a C3.2 flare observed by AIA
on 2010 August 1. It is found that the flare EUV emission at the foot-points exhibits a two-phase evolution
similar to the UV emission. We speculate that these EUV emissions are also generated in the 
lower atmosphere such as the transition region, which is impulsively heated and then 
cools down on coronal evolution timescales. This same notion was addressed
in a few previous studies. While modeling active region loops, \citet{Patsourakos2008, Klimchuk2009, Klimchuk2012} 
have shown that the transition region emission at the base of coronal loops contribute 
significantly to the total emission budget in EUV 171\AA\ such as observed by TRACE. 
Recently, \citet{Brosius2012} suggested that the simultaneous EUV emissions observed 
by AIA during the early phase of a B4.8 flare were produced by lower-atmosphere plasmas 
of a few hundred thousand degrees. On the other hand, some recent observations
by Soft X-ray Telescope \citep[SXT;][]{Tsuneta1991} \citep[e.g.][]{Mrozek2004} and by 
EIS \citep[e.g.][]{Milligan2011, Graham2013} have revealed high temperature emissions of up 
to 8 MK at the flare foot-points during the impulsive phase. In those events, hard X-ray emissions were also found at the
foot-points, and chromospheric evaporation is considered to be driven by precipitating non-thermal particles.
In this study, we will discuss the origin of the foot-point EUV emissions in this C3.2 flare
and their implication on flare modeling. The following section gives an overview of the flare, followed by observations
of the spatially resolved flare foot-point emissions in UV and EUV bands. In Section
4, we estimate UV and EUV emissions during the decay phase using a simple conductive
heating model, and conclusions and discussions are presented in the last section.

\section{Observations}
In this paper, we exclusively study the rise and decay of UV and EUV emissions in flare
{\em foot-points} identified from AIA observations of a C3.2 flare on 2010 August 1. 
A preliminary analysis of emissions from the {\em coronal loops} of the same flare observed
by AIA and GOES is presented in \citet{Qiu2012}. Figure 1 shows the light curves
of the total data counts summed in the active region in a few UV and EUV bands observed by AIA.
For clarity of presentation, in the plot, the minimum value is subtracted from each 
light curve, which is then normalized to its maximum. We note that whereas emissions 
in UV, soft X-ray, and EUV 94 band rise during the flare, the EUV 171
emission first decreases during the rise phase of the flare and then increases two hours later. 
The early attenuation of the EUV 171 emission is caused by disruption and disappearance 
of active region loops at the onset of the flare, which contributes to coronal dimming typically
observed in this wavelength \citep{Harra2007, Qiu2007}, \citep[also see][for recent observations]{Hock2013}. 

Shown in Figure 1, the flare is a long 
duration event with coronal radiation in soft X-ray and then subsequently EUV temperatures 
lasting for nearly four hours. Enhanced emissions at UV 1600\AA\ band lasts for two 
hours. Throughout the flare, UV or EUV flux observed by AIA is not saturated in any band, 
and the exposure time at any single band was a constant; therefore, the flare 
is a good candidate for quantitative analysis. RHESSI observations of this flare 
show gradual X-ray emission up to 20 keV similar to the GOES light curve, suggesting 
that the flare probably does not have significant non-thermal emissions. 

Flare emission in the UV 1600\AA\ broad-band is dominated by C{\sc iv} line 
emission, which is an optically thin line formed at 10$^5$K, the temperature 
of the upper chromosphere and transition region. Enhancement of this emission is observed at the 
feet of the flare loops, thereby forming the classic flare ribbons.  AIA also observes at the UV 1700\AA\ broadband, 
which mostly reflects the flare-enhanced UV continuum emission at the flare foot-points.
Past spectral observations suggest that UV continuum in these wavelengths is 
formed at the temperature minimum and thus characterized by temperatures of 4400 -- 4700 K in 
quiescent or active regions \citep{Brekke1994}.  Continuum enhancement during a flare is characterized by 
brightness temperatures up to 5400 K \citep{Cook1979}. \citet{Cook1979} also reported that the decay time of
this increased brightness temperature is comparable to the soft X-ray decay time. 
Given the large column depth of the temperature minimum region, 
these enhancements are not readily explained by direct heating from either thermal or non-thermal electron flux.  
Instead it is typically attributed to photo-ionization from short-wavelength emissions from above \citep{Machado1982, Phillips1992,
Doyle1992}.
This close causal link between the enhancements of C{\sc iv} and UV continuum explains the nearly identical morphology 
observed in the 1600\AA\ and 1700\AA\ images.

Assuming the continuum enhancement to be the same in both the 1700\AA\ and 1600\AA\ bands, although characterized 
by different regions of the black-body curve, we can use the former to remove the continuum from the latter. 
To do this we assume the 1700\AA\ band is dominated by the continuum emission \citep{Brekke1996} to estimate the 
brightness temperature of the enhancement in a given pixel.  We then subtract an amount from the 1600\AA\ band 
corresponding to the same brightness temperature.  The remainder, we contend, is an estimate of the C{\sc iv} emission from that pixel.

EUV emissions are usually produced in flare loops heated to a few tens of MK, 
and then cooled to 2-3 million K degrees or even below \citep[e.g.][]{Reale2012}. 
This general statement is supported by Figure 1, 
showing X-ray and EUV emissions characteristics of different temperatures peaking at different 
times. Figure 2 shows the time sequence of the flare evolution observed in UV 1600\AA\ broadband, 
as well as in five EUV bands at 304, 193, 335, 94, and 131\AA, which are roughly representative 
of increasing temperatures of coronal plasmas.
The figure shows the flare to consist of brightenings in two different loop systems.
A set of short loops in the north brightens first, followed by a set of long loops in the south.
For the same loop(s), emissions at relatively high temperatures (in 131 and 94 band, for example) occur
earlier than emissions at relatively low temperatures (in 193 and 304 bands, for example).

Apart from EUV emissions in flare loops, these images also show impulsive rise of EUV emission 
coincident with the UV emission at the same location during the early phase of the flare (left column of Figure 2).
These emissions arise where the flare loops, visible in EUV images minutes later, are rooted.
The origin of these emissions is the focus of this study. 

Images from the AIA multiple bands are rebinned to the scale of 1.2\arcsec\ by 1.2\arcsec, and 
are coaligned with each other with sub-arcsecond accuracy. Spatially resolved light curves, in units of data
number (DN) per second per pixel, are obtained in these bands. In the following analysis, we select 
the brightest UV foot-point pixels observed in 1600\AA\ that exhibit strong emission, or more
specifically those pixels with a count rate greater than five times the median count 
rate ($I_q = 71\ {\rm DN s^{-1}}$) of the quiescent region for more than three minutes. 
These pixels account for 50\% of all flaring pixels analyzed in \citet{Qiu2012}, but since these are
the brighter half, their total emission predominates the total UV emission of the flare.

\section{UV and EUV Emissions at the Foot-points of Flaring Loops}
\subsection{UV Emissions}
The top panels of Figure 3 show an example of the UV 1600 (dark dashed line in both panels) and 1700 
(dark solid line in the right panel) light curves from one of the brightest UV
pixels.  (This pixel is the brightest pixel within the small red box in the left column of Figure 2.) 
Most of the bright pixels exhibit a rapid rise for 5--10 minutes, 
followed by a gradual decay over a few tens of minutes. Such two-phase evolution 
was reported in stellar flares observed in a few UV lines including the C{\sc iv}
line \citep{Hawley2003}. Recently, \citet{Qiu2010, Qiu2012, Cheng2012, Liu2013} also reported
such an evolution pattern in UV 1600 emissions from spatially resolved flare kernels
observed by TRACE or AIA. So the two-phase evolution appears to be common in
UV emissions from flare foot-points. 

We also note that the observed 1700\AA\ emission exhibits a light curve
very similar to that of 1600\AA: an impulsive rise and gradual decay on timescales
identical to those observed in 1600\AA\ band.  While the morphology of the two light curves are identical, they are quantitatively quite different.  The 1600\AA\ emission is enhanced by an order of magnitude
over the pre-flare emission, while the peak 1700\AA\ emission is only about twice the pre-flare emission.  We attribute this difference to the contribution of C{\sc iv} to the former and not the latter.

The top panels of Figure 4 shows the UV light curves computed from the summed counts from 
all {\em foot-point} pixels identified in the UV 1600\AA\ band.  With all
flaring pixels summed up, emissions in the 1600\AA\ and 1700\AA\ bands rise above 
pre-flare levels by 150\% and 40\% respectively.

The broadband 1600\AA\ emission obtained by AIA includes contribution by the optically thin C{\sc iv}
line emission, which forms at the temperature of 10$^5$ K, the typical transition region temperature,
and the UV continuum forming around 4500 K degrees in the temperature minimum region. Both the line
emission and continuum emission are enhanced during the flare when the lower
atmosphere is heated. AIA also takes broadband images at UV 1700\AA\ with a few lines, whose net
contribution, however, may not dominate the emission in this broadband \citep{Brekke1996}. 
Comparison of images obtained in these two bands therefore help to distinguish C{\sc iv} emission from the UV continuum.

To the first order, we assume that the UV continuum in both bands is formed at the 
same brightness temperature $T_B$ described by Planck's function, and that the 
1700\AA\ emission is predominantly continuum emission. Taking into account the AIA 
instrument response function, the 1700\AA\ emission can then be used to compute the 
brightness temperature $T_B$. The red curve in the top right panel of
Figure 3 is the computed $T_B$ at the sample pixel during the flare. This temperature
varies from 4800 K before the flare to 5200 K at the peak of the flare, namely the brightness 
temperature is raised by 400 K for this bright flaring pixel. These numbers are within the reasonable range
in agreement with past UV spectroscopic observations of flares \citep{Cook1979}.


We then compute the continuum contribution to the 1600\AA\ band using the same $T_B$
and the response function of the AIA UV filter. The calculated 1600\AA\ continuum light curve
for that same pixel is shown as the blue curve in the top left panel in the figure, together with the
observed total count rate in this band, both in {\em absolute} scales. The comparison suggests
that whereas the pre-flare emission in this broadband is dominated by the continuum, during the flare,
the continuum emission contributes only a fraction of the total UV emission. The remainder UV emission
during the flare is most likely the contribution of the C{\sc iv} line (dark solid curve). 
In this bright pixel, the peak C{\sc iv} emission is about 4 times the continuum emission.
When summed over all flaring pixels (top left panel in Figure 4), the total C{\sc iv} emission 
(dark solid curve) is about 1.5 times the continuum emission (blue solid curve).

We caution that the above exercise gives an estimate of the possible contributions by the continuum
and the C{\sc iv} line emissions to the UV 1600\AA\ broadband. In this estimate, we have ignored contributions
by all other lines in both the 1600\AA\ and 1700\AA\ bands. On the other hand, by subtracting the
1700 emission off the 1600 band, contributions of these lines are partly cancelled. The estimate 
therefore only provides a first-order evaluation of C{\sc iv} emission in the flaring atmosphere. 

\subsection{EUV Emissions}
The other panels in Figure 3 show light curves (dark solid line) in 6 EUV bands for comparison to the UV 
1600\AA\ light curve (dark dashed line) from the same foot-point pixel. It is evident that EUV emission at 
one pixel typically exhibits at least two peaks, and that the first peak in each
EUV band is coincident with the UV emission peak.  Just like the UV light curve, the first EUV peak also exhibits 
a two-phase evolution, a rapid rise followed by a more gradual decay, and the rise and decay timescales
are entirely comparable with those of the UV light curves. 
The EUV filters of AIA are sensitive to plasmas with a range of temperatures including, in every case, 
a few hundred thousand degrees \citep{Lemen2012}. It is therefore very likely that
the first EUV peak is produced the same way UV emission is produced: impulsive energy deposition from
thermal conduction in the lower
atmosphere followed by a more gradual process correlated with plasma evolution in the overlying coronal loop.

The EUV emission, however, exhibits a more complicated structure than the UV light curve at the same foot-point pixel.
For example, in the 131 band, about ten to twenty minutes after the first peak, a second and more gradual 
emission peak shows up in the EUV light curve. In other EUV bands, the second peak occurs still later by up to 
two hours.  
While the first EUV peak occurs simultaneously in all EUV bands, i.e., independent of temperature, the timing
of the second EUV peak is wavelength dependent.  In general, emissions at EUV bands sensitive to higher temperatures
(e.g., 94, 131, and 335 bands with response function peaking at $>$3MK) tend to rise (when the first peak stops decaying) 
and peak earlier than the low-temperature sensitive bands (e.g., 211, 193, and 171 bands with response function
peaking at 1-2 MK). These observations convince us that the second-peak
EUV emission is explained by the standard picture of post-flare plasma cooling from ten to a few million K degrees.

Moreover, although the second EUV peak is observed in the same pixel
as the first peak, in most cases, the two peaks originate from plasmas in {\em different} parts
of {\em different} flare loops.  The first peak
is from the foot-point of a flare loop formed and heated earlier, and the second peak is a cumulative emission
by parts of the loops that are formed progressively and overlap on top of the foot-point of the earlier loop. 
Figure 2 confirms this scenario by comparing the morphology during the two peaks. It appears that, for the sampled
pixel, the first EUV peak occurs at the feet of the set of the short loops residing to the north west, and
the second EUV peak is rather associated with the set of the long loops tending to the south, and these
long loops in the south are formed and heated later than the short loops in the north \citep{Qiu2012}.
\citet{Woods2011} and \citet{Hock2013} suggest that, in many flares, these long loops associated with 
what they call the EUV late phase are related to the breakout model for CMEs.
 
Figure 4 shows the UV and EUV light curves of the total counts from all {\em foot-point} pixels identified
in the UV 1600\AA\ band. It appears that the total EUV light curves also exhibit two or more components.
The first component evolves the same way as the UV light curve independent of wavelength or temperature, 
and the second component evolves on timescales dependent on temperature. Again, the most likely scenario is that the
early phase EUV emissions from these pixels are indeed produced at the flare foot-points in the upper chromosphere
or transition region heated impulsively, and emissions later on are from later formed flare loops overlapping the 
foot-points brightened earlier. The second emission component, even if from a single pixel, is
a complex collection of coronal emissions from fractions of multiple loops that cannot be easily resolved. 
In the following text, we focus on discussing the two-phase evolution of the first peak occurred simultaneously in UV and EUV emissions.

\section{Foot-point UV/EUV Emission as a Coronal Pressure Gauge}
The two phases of the foot-point emission are governed by different physics.
The impulsive spike shown in the UV and EUV light curves is considered to be a signature of the lower 
atmosphere responding to energy deposition. It is most likely generated by a condensation shock
front propagating downward from the site of energy deposition by thermal conduction \citep{Fisher1989}.
The gradual decay, on the other hand, reflects the cooling of the overlying corona.
\citet{Hawley2003} reported such two-phase evolution in UV emissions from a few lines in stellar flare
observations, and found that these lines (including C{\sc iv}) can be used as a transition-region
pressure gauge monitoring evolution of coronal plasmas in overlying flare loops during the cooling
phase. During this phase, the entire loop is in approximate hydrostatic balance so the differential emission measure throughout the transition region is proportional to the equilibrium pressure --- the coronal pressure.  The emission from any line formed at transition region temperatures, such as C{\sc iv}, is therefore also proportional to coronal pressure.
In the following discussion, we explore whether this pressure-gauge logic can re-produce observed UV and EUV signatures.

\subsection{Transition Region Differential Emission Measure}
To find plasma evolution in overlying coronal loops, \citet{Qiu2012} used 
a zero-dimensional EBTEL model \citep{Klimchuk2008, Cargill2012} to 
calculate the mean temperature and density in the coronal loop. Inputs to the model include
the heating rate and loop length at each foot-point pixel.   
The latter is measured from the AIA imaging observations.  The
heating rate is inferred from the impulsive pulse of the UV light curve from that pixel, after using a single scaling parameter.
The EBTEL model also allows heat input either directly to the coronal plasma or non-thermal energy deposition in the lower atmosphere.  
As this particular flare exhibits very little non-thermal signature, we assume that the energy input was of the former variety.

The output of the EBTEL model are coronal plasma properties (mean temperature and density) 
which are used to compute the synthetic 
X-ray and EUV emissions by coronal loops observed by GOES and AIA. By matching the observed 
and synthetic X-ray and EUV fluxes, \citet{Qiu2012} were able to arrive at a first-order estimate 
to the scaling parameter used to convert impulsive emission to heating.
We note that in \citet{Qiu2012}, the earlier version of EBTEL model \citep[EBTEL1;][]{Klimchuk2008} was employed. 
In the present study, the coronal plasma properties are re-calculated using the updated version of EBTEL model 
\citep[EBTEL2;][]{Cargill2012}. The difference between the results by the two versions of the models is insignificant, mainly
because of the appropriate choice of free parameters guided by observations.

Figure 5 shows the mean temperature and density of the flux tube rooted at the bright 
pixel illustrated in Figure 3. The heating rate is inferred from the rise of the UV light curve
at this pixel with a duration of 10 minutes, and the best-guess magnitude of the heating flux is 
1.6$\times 10^9$ erg cm$^{-2}$ s$^{-1}$. The resultant pressure of the flux tube is plotted
in the middle panel, and it is seen that the decay of the UV 1600 light curve (and therefore the EUV 
light curves as well) evolves on the same timescale as the coronal pressure.  

This pressure is used to synthesize the C{\sc iv} emission following the ``pressure-gauge'' 
\citep{Fisher1987, Hawley1992}. We assume the transition region, where the spectral lines form, to be 
in hydrostatic balance at some pressure. The atmosphere is then structured by the balance 
between optically-thin radiative losses and conductive heat downward from the cooling coronal loop; 
plasma flow is neglected. With these conditions, the analytical solution is obtained to compute the differential
emission measure (DEM) along the leg of the flux tube \citep{Fisher1987, Griffiths1998, Hawley1992} to be
\begin{equation}
\xi_{se}(T) = \bar{P} \sqrt{\frac{\kappa_0}{8k_B^2}}T^{\frac{1}{2}}Q^{-\frac{1}{2}}(T) ~~.
\end{equation}
where
\begin{equation}
Q(T) = \int_{T_0}^{T} T'^{\frac{1}{2}}\Lambda(T')dT' ~~,
\end{equation}
and $\Lambda(T)$ is the optically-thin radiative loss function. Expressing the temperature dependent scaling constant 
as $g_{se}(T)$, we can compute the transition region DEM as $\xi_{se}(T) = g_{se}(T) \bar{P}$, which is directly proportional
to the mean pressure $\bar{P}$ computed using the zero-dimension EBTEL model \citep{Cargill2012}.

Plasmas inside flaring flux tubes are usually not in static equilibrium but undergo dynamic evolution.
During the heating phase, upflow (chromospheric evaporation) up to a few hundred kilometers per second 
is generated, and the decay phase is dominated by downflow (coronal condensation) of order
a few tens of km s$^{-1}$. Therefore, the transition region DEM should be corrected
with respect to flow terms; under the steady state assumption this is computed as \citep{Klimchuk2008}:
\begin{equation}
\xi_{ss}(T) = \bar{P} \frac{\kappa_0^{\frac{1}{2}}}{2k_B} [T^{\frac{1}{2}}\Lambda(T)]^{-\frac{1}{2}}(\sqrt{\gamma^2+1}+\gamma)
\end{equation}
where $\gamma$ is a function of mean coronal temperature $\bar{T_c}$ and flow speed $v_c$ across the coronal base, both calculated in EBTEL.
\begin{equation}
\gamma = \frac{5k_B T^{\frac{1}{2}}}{2\sqrt{\kappa_0 T^{\frac{1}{2}}\Lambda(T)}}\frac{-v_c}{\bar{T_c}}
\end{equation}
The above pressure gauge relation may be written as $\xi_{ss}(T) = g_{ss}(T) \bar{P}$.
Similar to Equation 1, the transition region DEM is scaled with the coronal pressure, the scaler $g_{ss}$
being dependent also on the plasma flow. For upflow, $v_c > 0$, and for downflow, $v_c < 0$.

The right panel in Figure 5 shows the transition region DEM $\xi_{se}$ and $\xi_{ss}$ in a few stages during 
the flux tube evolution. These few stages are indicated by the shaded bands in the left panel, representative 
of the rise, early decay, and late decay phase of the flux tube, respectively. The DEM in each stage is 
the mean value over 10 minutes. The DEM is modified when flow is included.

During the impulsive heating phase the loop is far from equilibrium and cannot be modeled in this manner.  
Rapid heating of the lower atmosphere from thermal conduction leads to upward and downward moving shocks 
\citep{MacNiece1986}. The upward shock is the leading edge of an evaporation flow of a hundred km/sec or more.  
The downward shock leads down flow, often called chromospheric condensation \citep{Fisher1989} in this 
initial phase of the flare, which is distinguished from the process of cooling and draining of coronal
plasma in the later phase. The effect, sometimes regarded as a rapid downward motion of the transition region, is to 
raise material at chromospheric densities to transition region temperatures. This results in emission 
from associated lines, such as C{\sc iv}, far in excess of what an equilibrium atmosphere might produce.  
This enhancement lasts only as long as the condensation shock does.  Afterward the loop assumes 
hydrostatic balance at its new pressure and begins cooling as described above.  
We attribute the brief, impulsive enhancement in C{\sc iv} to this scenario and use its 
amplitude and duration to infer the energy input into the coronal loop, but do not attempt to capture the 
physics in the EBTEL model.  Instead we quantify the relationship through a single empirical parameter, 
which we fix through observational comparison as described above.

\subsection{Comparison with Observations}
To compare with observations, we use the calculated transition region DEM to compute C{\sc iv} at the flaring pixel. 
The emissivity of the optically thin C{\sc iv} line $\epsilon(T)$ is
derived from CHIANTI 7.0 with ionization equilibrium \citep{Dere1997, Landi2012}. The total C{\sc iv} photon flux in units of photons 
cm$^{-2}$ s$^{-1}$ sr$^{-1}$ is computed using the DEM, and is converted to observed
count rate in units of DN s$^{-1}$ by convolving with the AIA instrument response 
function. Note that we have used the latest version of the response function released
in 2012 January, and with the correction factor from normalization to EVE observations; for
the 1600 band, this correction factor is 2.1, and for the 1700 band, this correction factor is
0.75 \citep{Boerner2012}.

This is then compared with the observed C{\sc iv} light curve. Comparison for a single pixel is given in the top
left panel in Figure 3, and the summed emission from all foot-point pixels is shown in the top left
panel in Figure 4. It is seen that, during the impulsive rise, the model calculated C{\sc iv} emission is far less than observed.
This is expected from the shock condensation scenario outlined in the previous section.  On the other hand, during the decay, 
the model calculated emission declines on the same timescale as observed; the amount of emission, 
computed with either static equilibrium or steady-state approximation is smaller than the observed flux by
a factor of 2 to 3 for the bright pixel. When emissions from all pixels are summed up, the computed
C{\sc iv} emission agrees with the observed total. This result indicates that the
pressure-gauge approximation can reproduce the observed decay timescale reasonably well; 
on the other hand, the magnitude comparison for single pixel and for all the pixels suggest that, the
observation or model or both of the C{\sc iv} emission seem to differ for differently heated flux tubes.

In the same way, we convolve the AIA instrument response functions of the six EUV bands with the computed
DEM to synthesize the EUV count rate light curves at the flare foot-points - the 304\AA\ band is not computed
since the formation mechanism of this line is more complex, for example, it is not optically thin. 
Figure 3 shows comparison of the EUV light curves for one pixel, and Figure 4 shows the sum of the emissions 
in all foot-point pixels. In these figures, the solid red curves and dashed red curves show the computed 
flux with static or steady-state approximations, respectively. In the later case, the computed flux 
is enhanced as coronal downflow into the transition region is included in the decay phase.

In synthesizing the EUV bands we integrate the DEM from 100,000 to 500,000 K only; we do not include the corona.
The upper bound of the temperature is rather arbitrary but not entirely unreasonable. The temperature 
distribution of the plasmas is along the length of the flare loop; however, we only look at one pixel 
at the foot-point. Because of the geometry of the loop on the solar disk, and the fact that flare loops in this 
event are very long with their half-length ranging from 50-100~Mm, only relatively cool 
plasmas at the bottom of the flux tube would contribute to emission at the foot-point pixel. 

The figures show that, again, the computed and observed EUV light curves for the first
peak decay on almost the same timescale, which is the decay timescale of the pressure in the flux tube.
In terms of magnitude, the computed emission is quite comparable with observed
in 335, 211, and 193 bands. The computed flux in 131 and 171 bands is higher by a factor of 3-4; on the other
hand, the computed flux in 94 band is smaller than observed by nearly a factor of 5. Note that \citet{Brosius2012} 
also conjectured that the low temperature response in the 94 band is likely under-estimated by a factor of 5.
With uncertainties in the effective upper-bound temperature that contributes to the foot-point emission, 
as well as in the low temperature response of AIA filters, it is still striking that the pressure gauge calculation
based on very simplified assumptions produces close estimates of the UV and EUV emissions at the flare
foot-points. 

The above experiments show that the pressure gauge calculation may be applied to the gradual cooling phase
when the flare loop is very close to equilibrium. On the other hand, the calculation
does not agree with the signatures during the impulsive heating phase, which is unlikely
to be in an equilibrium state. The static or steady-state equilibrium dictates that the plasma DEM
is proportional to the pressure which is uniform along the loop. Therefore, the ratio of optically thin
EUV or UV fluxes should remain a constant during its evolution. In Figure 6, we plot the ratio of the EUV flux
in a few bands to the C{\sc iv} flux as well as the ratio of EUV fluxes for the sample foot-point 
pixel during its evolution. It is shown that the flux ratio is almost a constant during the gradual
cooling phase, justifying the pressure-gauge assumptions. However, during the impulsive heating phase,
the flux ratio varies rapidly. Such behavior is observed in most foot-point pixels. It is therefore evident 
that the impulsive phase cannot be described by steady-state equilibrium. 

Finally, to estimate the contribution of coronal emission from flare loops on top of the
foot-point pixels, we plot the synthetic EUV emission by plasmas in flare loops \citep{Qiu2012}
but along a length of only 1 pixel. These are shown in the blue curves in the middle and bottom panels
in Figure 4. It is evident that in nearly every band the coronal emission is delayed with respect to the
foot-point emission. The peak of the coronal emission component is also temperature-dependent, with
high temperature emission (131 and 94 bands) peaking earlier than low temperature emissions (211, 193, 171 band).
During the first peak of the observed emission, the contribution by the coronal component is insignificant
except in the relatively hot bands, for example in the 131 band. 

\section{Conclusions and Discussions}
We have investigated the UV and EUV foot-point emissions observed by AIA during the early phase of the flare.
It is recognized that UV emission of the flare occurs at foot-points. We have shown that these same foot-points 
also produce EUV emissions observed by AIA, whose evolution is nearly identical to the UV light curve with a 
rapid rise on timescales of a few minutes followed by a gradual decay over a few tens of minutes in this
long duration flare. Therefore, these emissions are most likely produced by the same mechanism: impulsive 
heating of the lower atmosphere -- the upper chromosphere and transition region -- from a downward thermal 
conduction flux, and subsequent decay governed by the coronal plasma hydrodynamic evolution. 

Using a simple zero-dimensional loop heating model and loop heating rates empirically inferred from the rapid UV
pulse, we calculate mean properties of plasmas inside flaring loops, and in turn, compute the transition region
differential emission measure as scaled to the coronal pressure with static or steady-state approximations.
It is shown that the computed foot-point emissions in UV and EUV bands exhibit the same evolutionary timescale
as observed, which is the timescale of the coronal pressure. Assuming that the observed photon flux is produced
by plasmas at the coronal base with relatively low temperatures up to a few hundred thousand
K degress, the amount of computed emission compares well with observed in 7 bands by within a factor of 3-5, 
a fairly good agreement given uncertainties in the loop geometry and the AIA response functions at low temperatures. 
This simple exercise suggests that evolution of flare foot-point emissions may be used to 
monitor coronal plasma evolution, and shows the importance of coupling the coronal and lower atmosphere heating 
and dynamics as independent constraints to loop heating models.

It is noted that the transition region DEM may be substantially increased at temperatures beyond a few hundred thousand
K degrees. Computed EUV flux taking into account these higher temperature plasmas, however, produce a lot 
more flux than observed by one to two orders of magnitude. This may indicate either a temperature 
dependent filling factor of this order, or that the static or steady-state assumptions are not a good 
approximation for plasmas at higher temperatures.

The above experiment does not reproduce the impulsive pulse of UV and EUV emissions in the first few minutes, indicating
that steady-state assumptions and/or assumed equilibrium conditions used to calculate UV and EUV lines in CHIANTI
are not adequate for this period of impulsive heating. It is also plausible that non-thermal particles
heat the lower atmosphere during this phase; however, there is no strong evidence for the presence of
these particles in this event. A more sophisticated hydrodynamic modeling, aided with imaging 
spectroscopic observations of the flare foot-points will help gain insight in this phase. 


\acknowledgments We thank Dr. R. C. Canfield for illuminating us about the flare energetics and lower-atmosphere heating.
We thank the referee for careful reading of the manuscript and constructive comments that help improve clarity
of the paper. We acknowledge SDO for providing quality observations. The work by JQ and WL is supported by NSF grant ATM-0748428. 
Part of the work was conducted during the NSF REU Program at Montana State University. DWL acknowledges support
by the NASA Supporting Research and Technology Program. The work of JAK was supported by the NASA Supporting 
Research and Technology Program.

\bibliography{local}

\begin{figure}
\epsscale{0.7}
\plotone{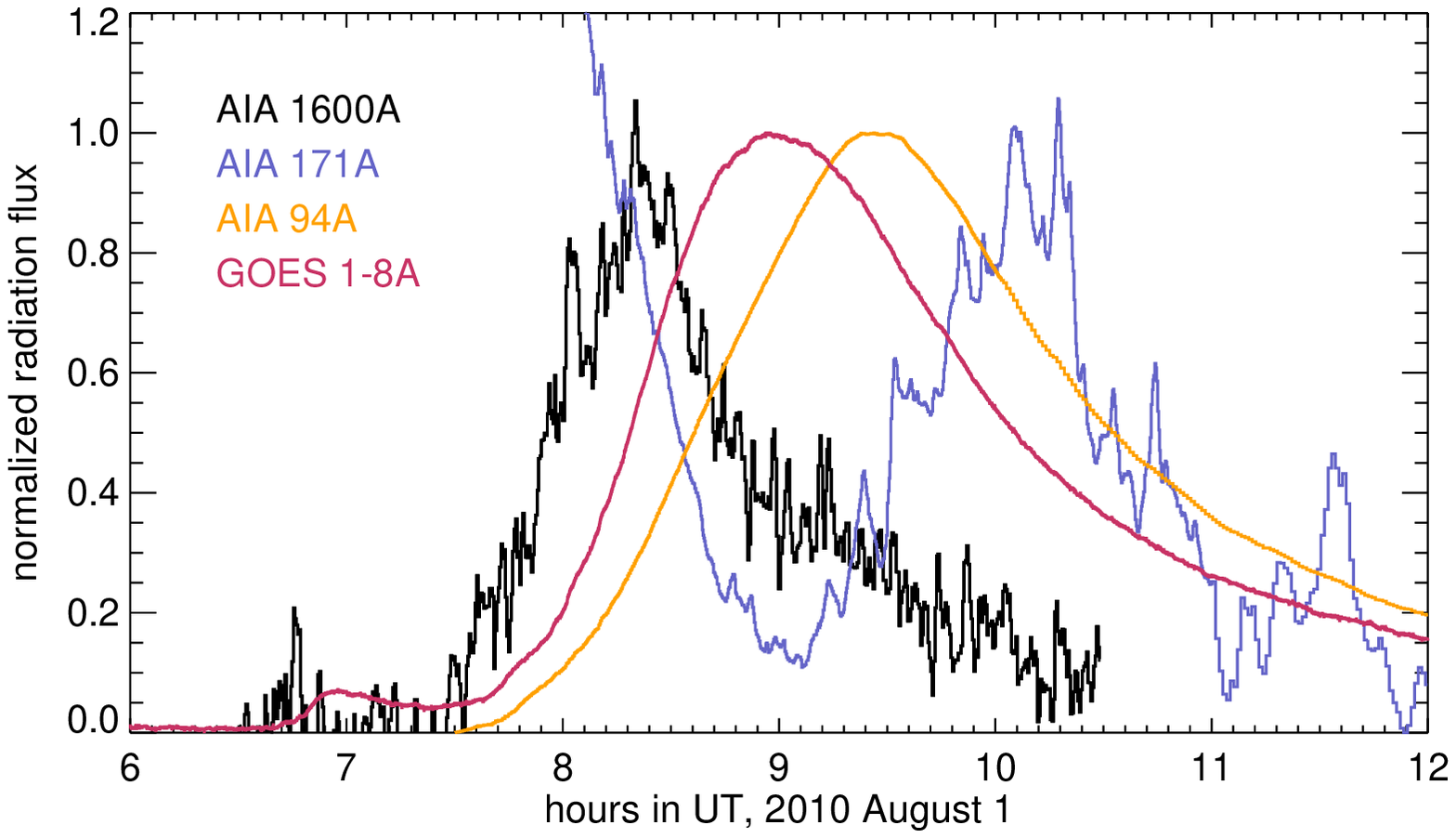}
\caption{Light curves of the 2010 August 1 C3.2 flare in UV 1600\AA\ and EUV 171\AA\ and 94\AA\ by SDO/AIA,
and soft X-ray 1-8\AA\ by GOES.} \label{fig:overview}
\end{figure}

\begin{figure}
\epsscale{1.0}
\plotone{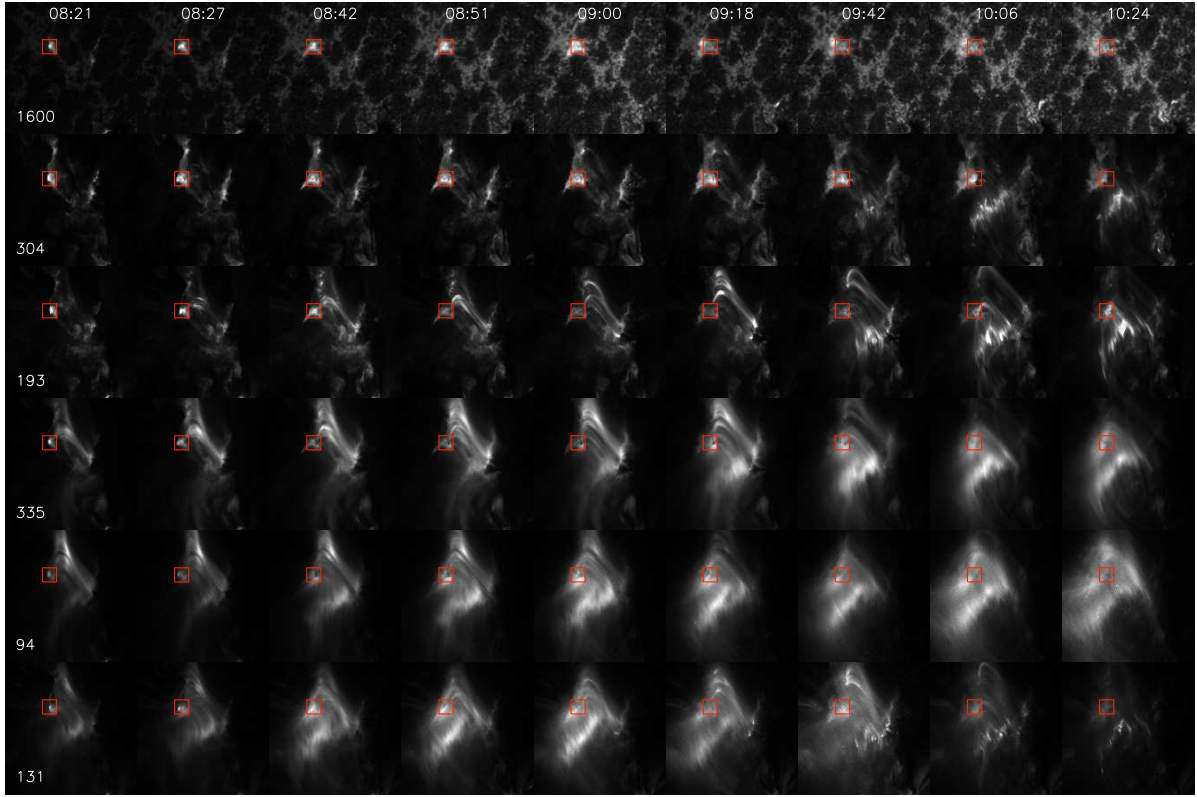}
\caption{Evolution of the flare as observed in AIA UV 1600\AA\ band and EUV 94, 131, 193, 304, 335 bands. Images at EUV 
171 and 211\AA\ are not shown, since the flare morphology in these two bands is similar to that observed in 193\AA\ band.
The red box in the figures show the location of the sample foot-point pixel, which is impulsively brightened and 
identified in the UV 1600\AA\ images. Images of the left column show UV and EUV images at the time when this pixel 
is brightest; it is seen that this same pixel is brightest at all bands. Images in other columns show time 
evolution after the impulsive brightening at this pixel. Whereas UV 1600\AA\ images only exhibit emission 
at the foot-point, all EUV images show flare loops connected at or overlapping upon this sample foot-point pixel. The
times of these images are also indicated by dotted lines in Figure 3. } \label{fig:risedecay}
\end{figure}

\begin{figure}
\plotone{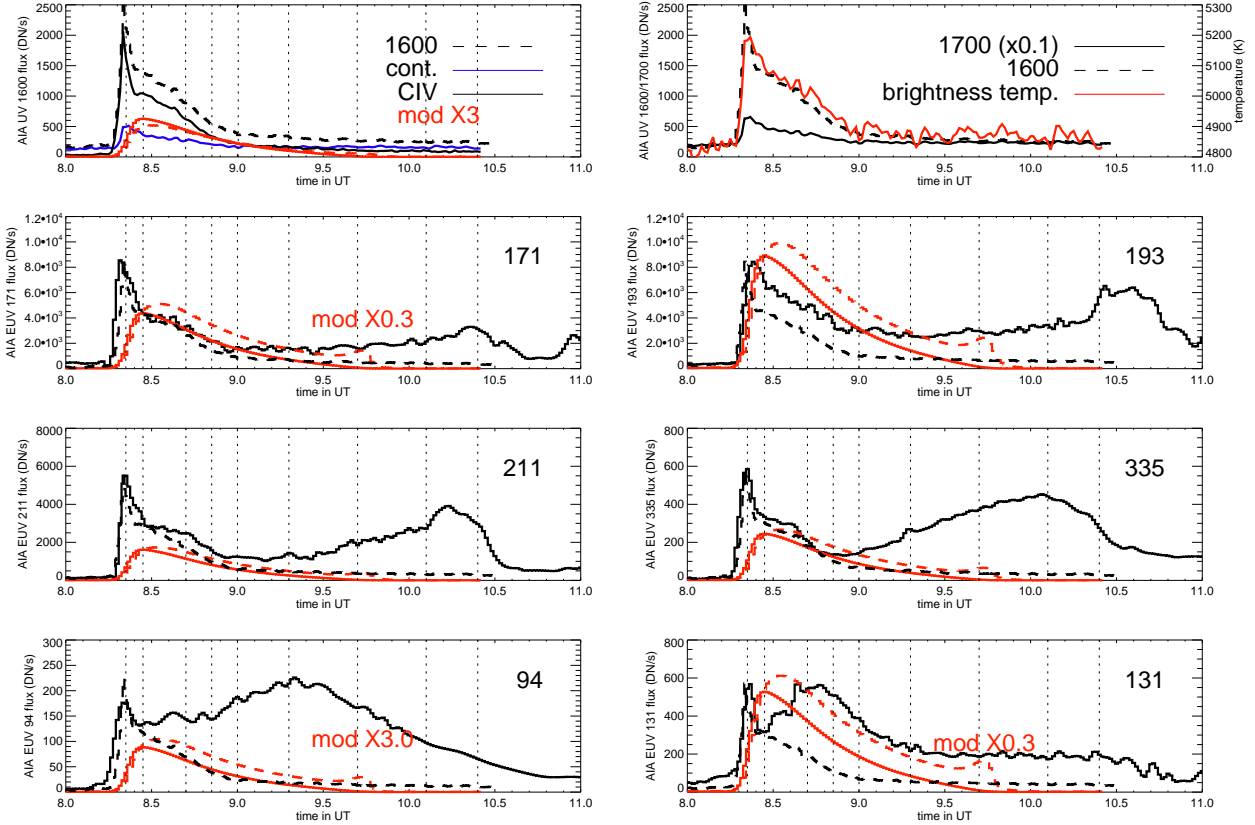}
\caption{Top: observed UV 1600 (dashed black in both panels) and 1700 (solid back in the right panel) 
light curves of the sample foot-point pixel, superimposed with the estimated continuum light curve in
the 1600 band computed using the brightness temperature (red in the right panel), 
and C{\sc iv} emission light curve, and compared with model computed UV C{\sc iv} light curve (red in the left panel).
Middle and bottom: observed EUV count rate light curves at the sample foot-point pixel, superimposed
with model-calculated light curve under static equilibrium (solid red) and steady-state (dashed red) approximations, respectively.
Note that the model computed light curves in the 171, 94, and 131 bands are multipled by factors of 0.3, 3, and 0.3, respectively.
In all the EUV panels, the black dashed curve shows the observed UV 1600\AA\ light curve arbitrarily scaled.
The vertical dotted lines indicate the times of the snapshot images in Figure 2.
} \label{fig:one}
\end{figure}

\begin{figure}
\plotone{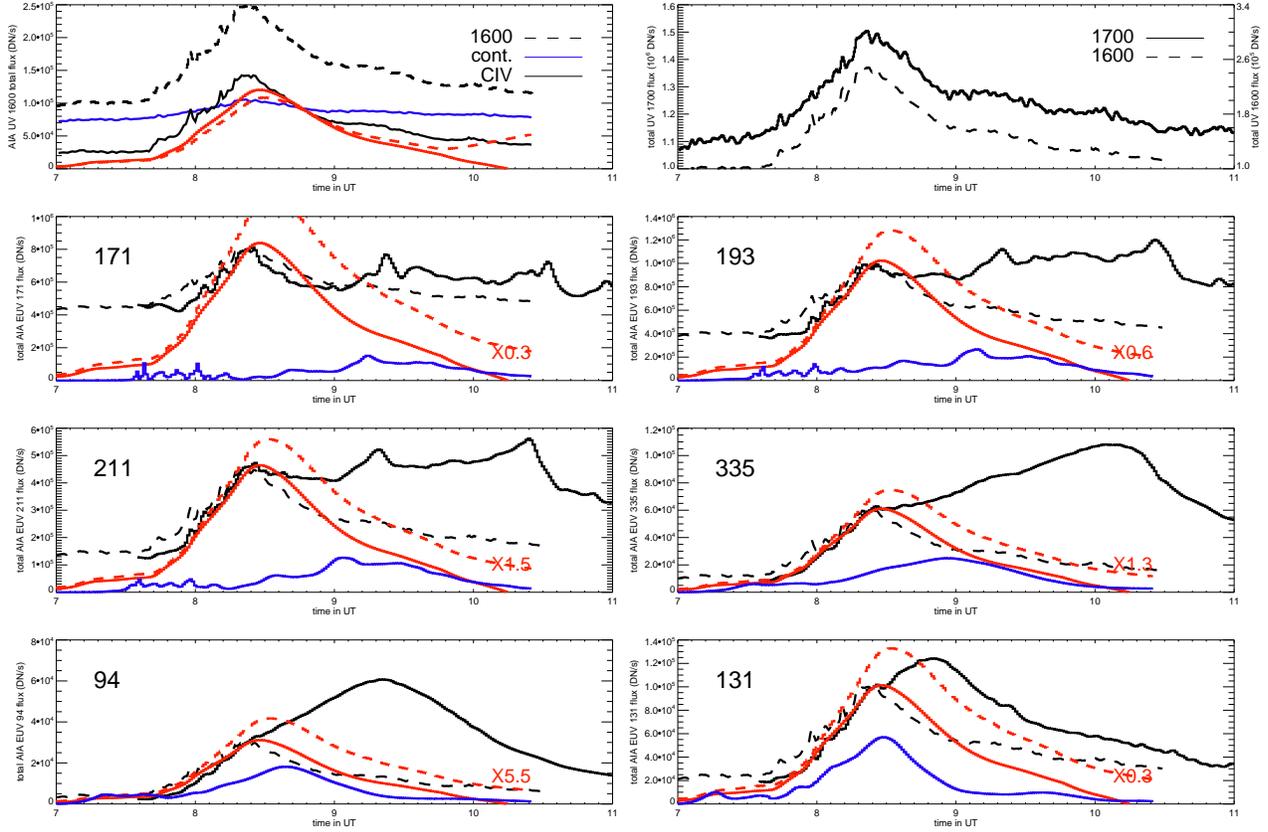}
\caption{Observed UV and EUV light curves (solid black) summed from all foot-point pixels identified from the
UV 1600\AA\ images, compared with model computed UV C{\sc iv} light curve and EUV light curves with static 
equilibrium (solid red) and steady-state (dashed red) approximations, respectively.
Note that the model computed EUV light curves are multipled by factors indicated in the figure.
In all the EUV panels, the black dashed line shows the observed UV 1600\AA\ light curve arbitrarily scaled.
} \label{fig:total}
\end{figure}

\begin{figure}
\plotone{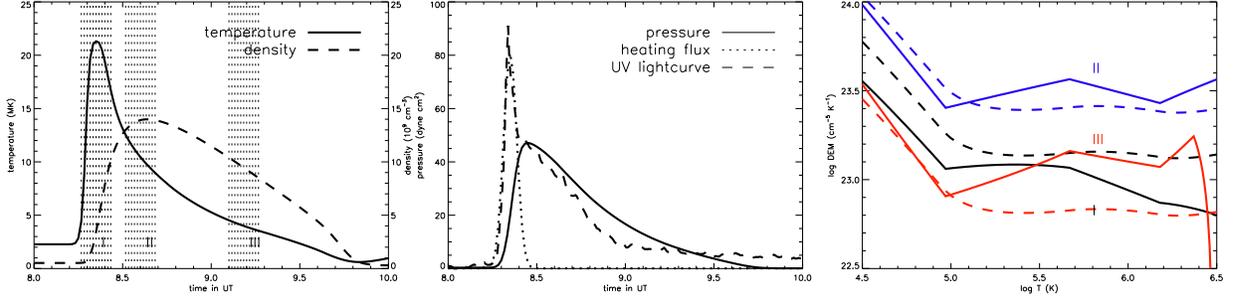}
\caption{Left: mean temperature (solid) and density (dashed) of the half loop rooted at the sample foot-point pixel; Middle:
the UV 1600\AA\ light curve at the foot-point pixel (dashed), the inferred loop heating rate (dotted)
from the rise of the UV light curve, and the mean pressure (solid) of the coronal plasma. The temperature, density, and pressure
are computed using EBTEL-2 model \citep{Cargill2012}. Right: the mean transition region DEM of this loop
averaged over 10 minutes in different stages of the flux tube evolution indicated by the shaded bands in the left panel. 
Solid lines show the DEM $\xi_{se}$ computed with the static equilibrium assumption, and dashed lines show the DEM $\xi_{ss}$
with the steady-state assumption.
} \label{fig:loop}
\end{figure}

\begin{figure}
\plotone{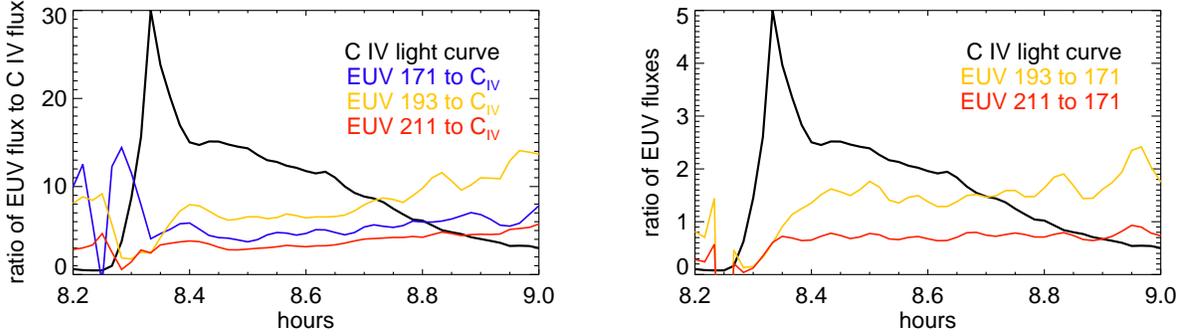}
\caption{Left: ratio of observed EUV fluxes in 171, 193, and 211 bands to the C{\sc IV}
flux for the sample pixel during the flare. Right: ratio of observed 
EUV fluxes in 193 and 211 bands to that in 171 band for the same pixel. In both panels,
the C{\sc IV} light curve, arbitrarily scaled, is plotted to provide information of
the evolution of the foot-point emission.} \label{fig:loop}
\end{figure}

\end{document}